\documentclass[useAMS,usenatbib]{mn2e}
\usepackage{wrapfig}
\usepackage{deluxetable}
\newcommand\chandra{{\it Chandra}}

\newcommand\xmm{{\it XMM-Newton}}
\newcommand\swift{{\it Swift}}

\newcommand\Msun{\ifmmode M_{\odot} \else $M_{\odot}$\fi}

\newcommand\kms{\ifmmode {\rm~km\ s}^{-1} \else ~km s$^{-1}$\fi}
\newcommand\Hunit{\ifmmode {\rm~km\ s}^{-1}\ {\rm Mpc}^{-1}
        \else ~km s$^{-1}$ Mpc$^{-1}$\fi}
\newcommand\ctssec{\ifmmode {\rm~count\ s}^{-1} \else ~count s$^{-1}$\fi}
\newcommand\ergsec{\ifmmode {\rm~erg\ s}^{-1} \else
        ~erg s$^{-1}$\fi}
\newcommand\funit{\ifmmode {\rm~erg\ s}^{-1}\;{\rm cm}^{-2} \else
        ~ergs s$^{-1}$ cm$^{-2}$\fi}
\newcommand\phflux{\ifmmode {\rm~photon\ s}^{-1}\;{\rm cm}^{-2}
        \else   ~photon s$^{-1}$ cm$^{-2}$\fi}
\newcommand\efluxA{\ifmmode {\rm~erg\ s}^{-1}\;{\rm cm}^{-2}\;{\rm
        \AA}^{-1} \else ~erg s$^{-1}$ cm$^{-2}$ \AA$^{-1}$\fi}
\newcommand\efluxHz{\ifmmode {\rm~erg\ s}^{-1}\;{\rm cm}^{-2}\;{\rm
        Hz}^{-1} \else ~erg s$^{-1}$ cm$^{-2}$ Hz$^{-1}$\fi}
\newcommand\cc{\ifmmode {\rm~cm}^{-3} \else cm$^{-3}$\fi}
\newcommand\FWHM{\ifmmode {\rm~FWHM} \else ${\rm~FWHM}$\fi}
\newcommand\Lsun{\ifmmode L_{\odot} \else $L_{\odot}$\fi}

\newcommand\hbeta{\ifmmode {\rm H}\beta \else H$\beta$\fi}
\newcommand\Kalpha{\ifmmode {\rm K}\alpha \else K$\alpha$\fi}
\newcommand\nh{\ifmmode N_{\rm H} \else N$_{\rm H}$\fi}
\usepackage{graphicx}
\usepackage{color}

\newcommand\mnras{{MNRAS}}
\newcommand\apj{{ApJ}}

\newcommand\apjs{{ApJS}}
\newcommand\aj{{AJ}}
\newcommand\araa{{ARA\&A}}
\newcommand\aap{{A\&A}}
\newcommand\apss{{Ap\&SS}}

\usepackage{graphicx}
\usepackage{color}

\title{X-ray outbursts from a new transient in NGC 55}

\author[V. Jithesh et al.]{V. Jithesh\thanks{E-mail: jithesh@shao.ac.cn} and Zhongxiang Wang \\
Shanghai Astronomical Observatory, Chinese Academy of Sciences, 80 Nandan Road, Shanghai 200030, China}

\begin{document}
\maketitle
\label{firstpage}

\begin{abstract}
We report the outbursts from a newly discovered X-ray transient in 
the Magellanic-type, SB(s)m galaxy NGC 55. The transient source, 
XMMU J001446.81-391123.48, was undetectable in the 2001 \xmm{} and 
2004 \chandra{} observations, but detected in a 2010 \xmm{} observation 
at a significance level of $9\sigma$ in the 0.3--8 keV energy band. 
The \xmm{} spectrum is consistent with a power law with photon index 
$\Gamma = 3.17^{+0.22}_{-0.20}$, but is better fit with a 
$kT_{in} = 0.70\pm0.06$\,keV disk blackbody. The luminosity was
$\sim 10^{38}$\,erg\,s$^{-1}$, and the source displayed strong short-term X-ray 
variability. These results, combined with the hardness ratios of its emission, 
strongly suggest an X-ray binary nature for the source. 
The follow-up studies with \swift{} XRT observations revealed that the source 
exhibited recurrent outbursts with period about a month. The XRT spectra can 
be described by a power law ($\Gamma\sim 2.5$--2.9) or
a disk blackbody ($kT_{in}\sim 0.8$--1.0\,keV), and the luminosity was
in a range of 10$^{38}$--10$^{39}$\,erg\,s$^{-1}$, with no evidence showing
any significant changes of the spectral parameters in the observations.
Based on the X-ray spectral and temporal properties, we conclude 
that XMMU J001446.81-391123.48 is a new transient X-ray binary in NGC 55, which 
possibly contains a black hole primary.
\end{abstract}



\section{Introduction}

Transient X-ray sources have been extensively studied from the beginning of 
X-ray astronomy. Most of them are binary systems with a black hole (BH) or a
neutron star (NS) as the primary. 
These systems have been primarily discovered when they entered outbursts 
characterised by an episode of high accretion rates and abrupt increases
of X-ray luminosity by several orders of magnitude. During an outburst, 
the source goes through different spectral states: {\it low/hard state} (LHS) 
and {\it high/soft state} (HSS) are the two principal states of black hole 
X-ray binaries (BHXBs). In the LHS, the spectrum is described by a hard power 
law with photon index in the range of 1.5 - 2.0, and in the latter, 
thermal emission ($kT \sim$ 1 keV) from an optically thick and geometrically 
thin accretion disc dominates. There is also a {\it steep power law state} (SPL), 
described by a disk blackbody component plus power law with photon index 
$\Gamma > 2.4$ \citep[see][for extensive reviews on spectral states]{Mcc06, Rem06}. 
The outbursts typically last a few months with recurrence 
period of many years \citep[$\sim \rm 2-57.8~yr$;][]{Che97, Kuu97}. 
NS X-ray binaries also have such outbursts with much shorter recurrence 
period \citep[$\sim 100-200$ d;][]{Gue99, Mas02} compared to the BH systems. 
However, BH and NS systems both show many remarkable similarities in their 
spectral states \citep{Van94, Cam98, Mcc06}.

NGC 55 is a Magellanic-type, SB(s)m galaxy and its X-ray properties has 
been well studied by two \xmm{} observations conducted in 2001 \citep{Sto06}. They identified 
137 X-ray sources in the field of view and classified 42 X-ray sources 
in the $D_{25}$ region of the galaxy as X-ray binaries (XRBs), supernova 
remnants, and very soft sources. Moreover, this galaxy hosts a very bright 
BHXB candidate, which showed a marked upward drift, significant chaotic 
variability and pronounced dips in the \xmm{} observations \citep{Sto04}. 
This particular source belongs to the ultraluminous X-ray source class with 
X-ray luminosity $> 10^{39}~\rm erg~s^{-1}$. 

In this paper, we report the outbursts from a new X-ray transient in NGC 55. 
The transient source was discovered serendipitously from the inspection of 
archival {\it XMM-Newton} observations, which were primarily selected 
for studying the variability of super soft X-ray sources. While a more 
detailed report 
on the variability of super soft X-ray sources will be presented later, 
in this paper, we concentrate on this transient which exhibited
recurrent outbursts in the \swift{} X-ray Telescope (XRT) observations. 
In the following, \S 2 describes the observations and data reduction 
techniques used. We explain the analysis and results in \S 3 and 
discuss the results and possible nature of the transient in \S 4. 

\section{Observations and Data Reduction}

We used archival {\it XMM-Newton} observations of NGC 55 performed in 2001 
November and 2010 May. The data from the European Photon Imaging Camera (EPIC) 
PN detector were reduced and analysed using standard tools of the 
{\it XMM-Newton} Science Analysis Software (SAS) version 13.5.0. 
The full-field background light curve was extracted from the PN camera 
to remove the particle flaring background and create the good time intervals 
file. A count rate of $\ge~0.8~\rm ct~s^{-1}$ in the 10--12 keV light 
curve was used for the rejection. We used the PN events with the best quality 
data (FLAG = 0) and PATTERN $ \le 4$, and removed the hot pixels in the data by using the flag 
expression $\#XMMEA\_EP$. The source detection routine ({\tt EDETECT\_CHAIN}) 
was carried out using the standard parameters for EPIC-PN data over 
the entire energy bands and obtaining the final source list from 2001 and 2010 
observations. We corrected the X-ray source positions by correlating 
the final source list with the USNO A2.0 optical catalogue \citep{Mon98}, 
using the SAS task {\tt EPOSCORR}.

We also analysed two archival {\it Chandra} observations conducted in 2001 
and 2004. The {\it Chandra} data were reduced and reprocessed using 
the science threads of {\it Chandra} Interactive Analysis of Observations 
(CIAO) version 4.6 and HEASOFT version 6.15.1. In addition, we 
used {\it Swift} XRT observations of the region carried out during 2013 
April to 2014 October. We selected the observations with an exposure time 
of $> 2~\rm ks$ from the {\it Swift} program, which resulted 24 
observations (see Table \ref{log}). We reduced the {\it Swift} data using 
HEASOFT and the Calibration Database (CALDB) files as of 2014 June 10. 
We processed the photon counting mode of {\it Swift} observations using 
the {\it Swift} specific {\tt FTOOL XRTPIPELINE}, by following the standard 
procedures.

\begin{table}
\tabletypesize{\footnotesize}
\tablecolumns{4}
\tablewidth{240pt}
\setlength{\tabcolsep}{2pt}
  \caption{Observation Log}
  \begin{tabular}{@{}cccc@{}}
  \hline
\colhead{Mission}&\colhead{ObsID}&\colhead{Date}&\colhead{Exposure\tablenotemark{a}} \\

\hline 
{\it XMM-} & 0028740201, 0028740101 & 2001 Nov & 33.6, 31.5 \\
{\it Newton} & 0655050101 & 2010 May & 127.4 \\
{\it Chandra} & 2255 & 2001 Sep & 60.1 \\
	      & 4744 & 2004 Jun & 9.7 \\

\swift{} & K01 - K07 & 2013 Apr - May & 4.9 - 5.6 \\
	 & K09 - K20 & 2013 Jun - Aug & 4.5 - 4.7 \\

	 & L01, L04 & 2013 Sep - Nov & 3.5, 2.7 \\

	 & M01, M03, M05 & 2014 Oct & 2.9, 2.2, 3.7 \\
\hline

\end{tabular} 
\tablecomments {The prefix K, L and M on \swift{} denotes 000326190, 000821200 and 000334680 respectively.}
\tablenotetext{a}{Exposure time is in units of kilo-seconds.}
\label{log}
\end{table}

\section{Analysis and Results}

\subsection{Source Identification}

\begin{figure}
\centering
\includegraphics[width=4.1cm,height=4.1cm]{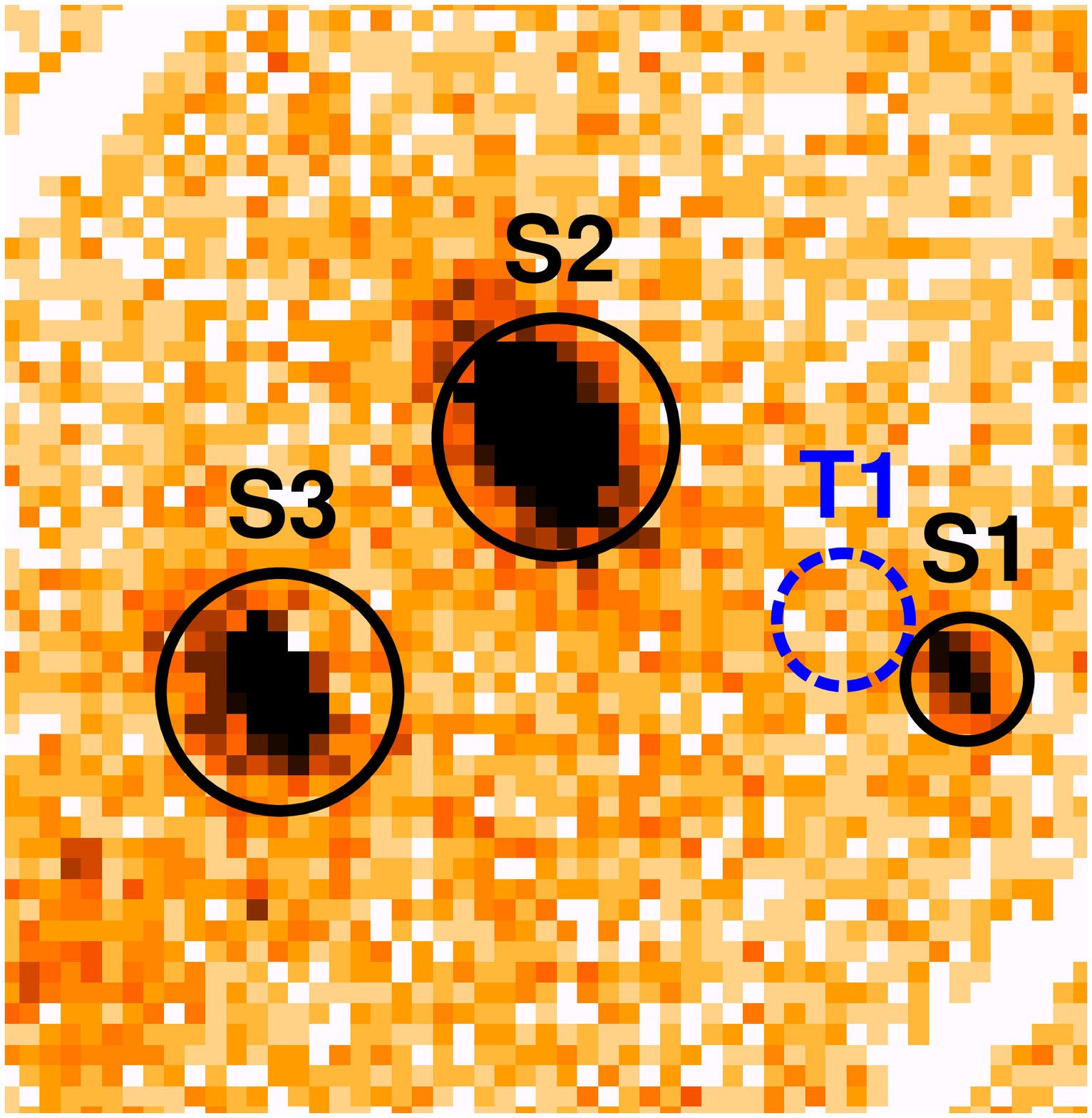}
\includegraphics[width=4.1cm,height=4.1cm]{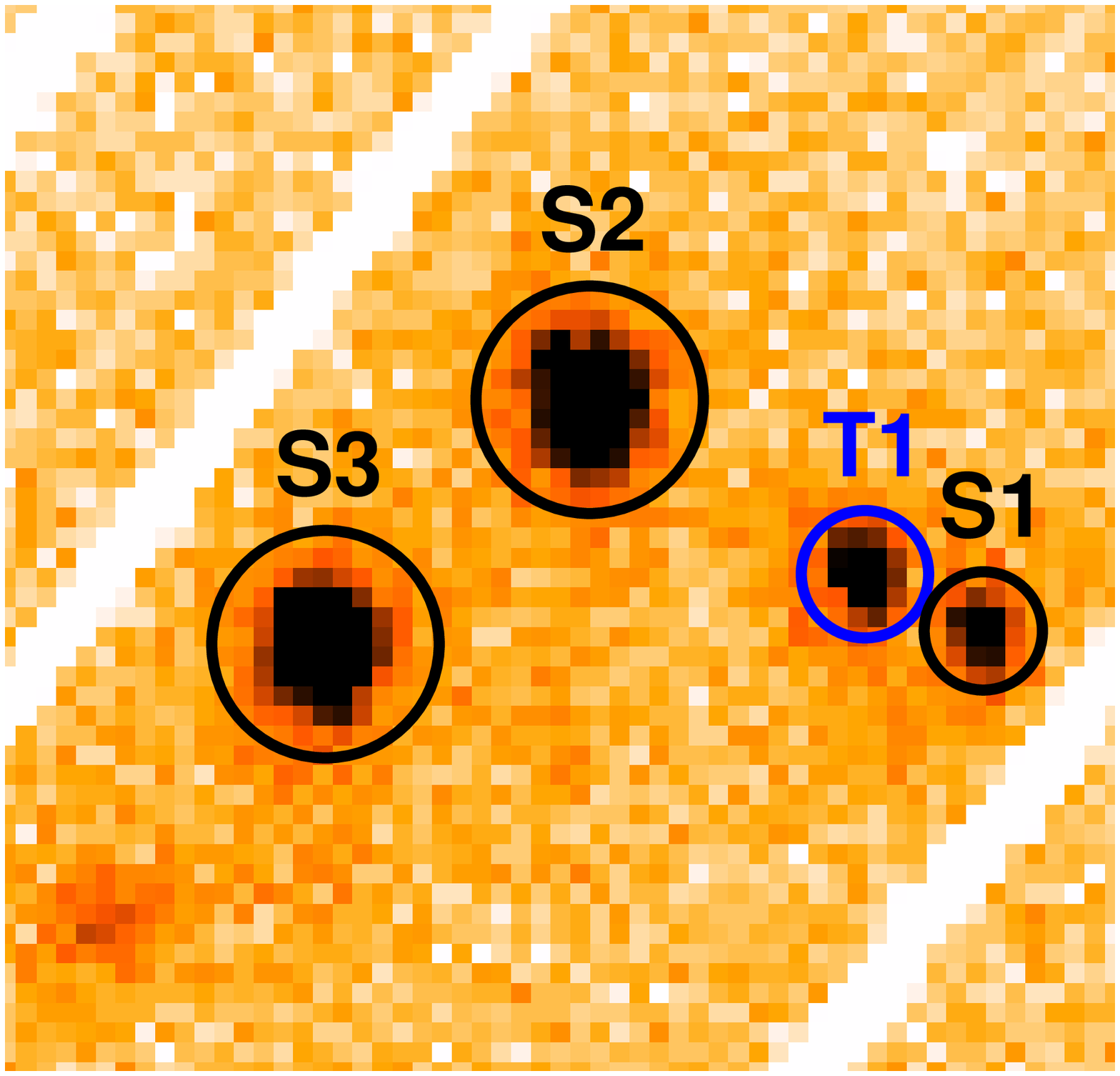}
\caption{A 4 arcmin~$\times$ 4 arcmin~ portion of the EPIC-PN image
including the transient source (T1) in the 2001 (left; ObsID: 0028740201) and 2010 
(right; ObsID: 0655050101) {\it XMM-Newton} observations respectively. 
Three bright sources (S1, S2 and S3) were detected in both observations. 
The transient source, XMMU J001446.81-391123.48, was detected
only in the 2010 observation.}
\label{pos}
\end{figure}

The new transient was discovered by an inspection of the image from 
the 127 ks {\it XMM-Newton} observation conducted in 2010. To determine 
the source position, we compared the astrometrically corrected X-ray 
positions of the persistent sources in 2001 and 2010 observations. 
In Fig. \ref{pos}, we show the field of three bright sources (S1, S2 and S3), 
which were identified in the catalogue of \cite{Sto06} as 
source\#38 (XMMU J001444.62-391135.9), 43 (XMMU J001452.02-391045.2), and 
47 (XMMU J001457.00-391139.2), respectively. 
We also detected a new source, XMMU J001446.81-391123.48, only in 
the 2010 observation with positional uncertainty of 0.5 arcsec ($1\sigma$ significance).
This source is located within the $D_{25}$ ellipse of NGC 55. 

We searched the on-line catalogues for X-ray sources consistent with 
the position of XMMU J001446.81-391123.48, but did not find any likely 
candidates. The analysis of {\it Chandra} observations conducted in 2001 
and 2004 did not find any sources in the XMMU J001446.81-391123.48 positional
error region. Also this particular source was not catalogued in 
the {\it XMM-Newton} observations \citep{Sto06} and previous {\it ROSAT} 
observations \citep{Rea97, Sch97}, but only in the third generation \xmm{} 
Serendipitous Source Catalogue (3XMM-DR4)\footnote{http://xmmssc-www.star.le.ac.uk/Catalogue/3XMM-DR4/}.
Based on these, we concluded that XMMU J001446.81-391123.48 is a previously 
undetected source and it is a new transient in NGC 55. 

To determine the significance of detection, we extracted the 0.3--8 keV counts 
from a 14 arcsec radius circle centred at position of the source 
(R.A.=$0^{\rm h}14^{\rm m}46^{\rm s}.81$, DEC=$-39^{\circ}11'23''.48$, 
equinox J2000.0). The background level was determined using a nearby 
source-free circular region with a radius the same as that for the source. 
After background subtraction, we obtained $1413\pm42$ and $721\pm29$ counts 
from EPIC-PN and MOS respectively; the combined detection significance 
is $9.4\sigma$. 

\subsection{\xmm{} and {\it Chandra} Data Analysis}

The hardness ratios (HRs) were calculated from the count rates and 
defined as HR1=(M$-$S)/(M+S) and HR2=(H$-$M)/(H+M), where S, M and H are 
the count rates in soft (0.3--1 keV), medium (1--2 keV) and hard (2--6 keV) 
bands respectively. The ratios obtained for XMMU J001446.81-391123.48 
are HR1=$0.45\pm0.01$ and HR2=$-0.38\pm0.01$. We used the X-ray colour 
classification scheme, tuned for {\it XMM-Newton} data \citep{Jen05}, 
to classify the source. Using the same colour criteria as employed 
by \cite{Jen05}, we classified this source as an X-ray binary. Although we 
cannot definitely classify any sources by their X-ray colours alone, 
but this approach is a first step to identify the class for the source.   

\begin{figure}
\centering
\includegraphics[width=8.0cm,height=6.5cm,angle=0]{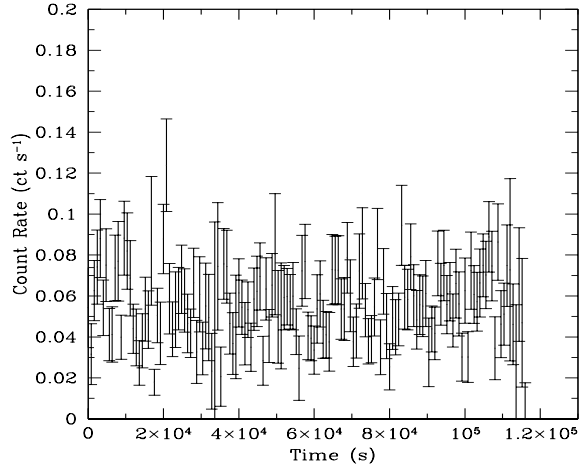}
\caption{Combined EPIC-PN and MOS 0.3--8 kev light curve of XMMU J001446.81-391123.48. The light curve has been background subtracted and with 800 s binning.}
\label{light}
\end{figure}

We extracted the background subtracted light curve for the transient source 
based on the combined EPIC-PN and MOS camera over 0.3--8 keV energy range. 
The resulting light curve, binned with 800-s, is shown in Fig. \ref{light}.
The short-term X-ray variability of the source was investigated by carrying 
out a Kolmogorov--Smirnov (K-S) test on an 800-s binned light curve and the 
additionally extracted 100-s binned light curve. From the K-S test, we found
that the source showed a strong short-term X-ray variability at confidence 
level of $> 99.99\%$ in both light curves.      

\begin{figure}
\centering
\includegraphics[width=6.5cm,height=7.5cm,angle=-90]{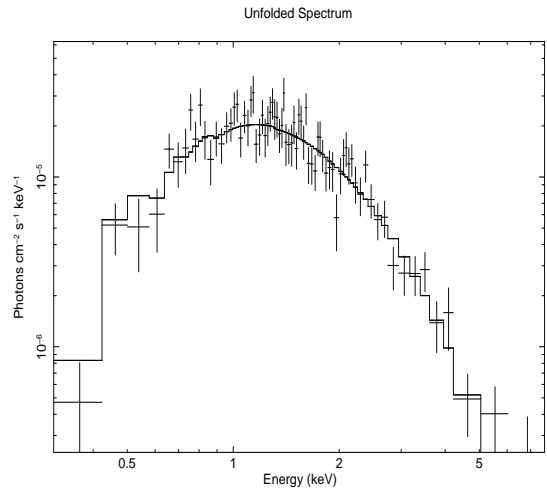}
\caption{The 0.3--8 keV \xmm{} EPIC-PN spectrum of the transient source, fitted with an absorbed disk blackbody model.}
\label{spectrum}
\end{figure}

The source and background spectra, together with response and ancillary 
response files, were obtained using the standard SAS tasks. 
The source spectrum grouped to a minimum of 20 counts per bin and 
the spectral analysis was performed with {\tt XSPEC} version 12.8.1g. 
The X-ray luminosity was calculated by assuming the distance of 
1.78\,Mpc \citep{Kar03}. The source spectrum was fitted with single-component 
models, power law ({\tt PL}), multi-colour disk blackbody ({\tt DISKBB}), 
and blackbody ({\tt BBODY}). An absorption component ({\tt TBABS}) was also
added to each model. The spectrum is best described by the {\tt DISKBB} model 
with $kT_{in}=0.70\pm0.06~\rm keV$, $\chi^2/d.o.f = 70.5/71$, while the power 
law model, with photon index $\Gamma=3.17^{+0.22}_{-0.20}$, $\chi^2/d.o.f = 74.2/71$, 
also provides a statistically acceptable fit. In Fig. \ref{spectrum}, 
the spectral fit with the {\tt DISKBB} model is shown.
However, spectral fit became worse (statistically) with {\tt BBODY} model, 
$\chi^2/d.o.f = 80.1/71$, compared to the power law and disk blackbody 
models. The unabsorbed luminosity for the power law and disk blackbody 
spectral fits are, $2.40^{+0.84}_{-0.54} \times 10^{38}$ and 
$5.75^{+0.42}_{-0.38} \times 10^{37}~\rm erg~s^{-1}$ respectively. The intensity 
for the power law model increases indefinitely as the energy decreases, 
while the disk blackbody intensity decreases towards the lower energies. 
Thus, there are substantial differences in line-of-sight absorption 
and luminosity as we see in our spectral fit. 
These results are summarized in Table \ref{fit}.  
We note that the column densities obtained are higher than 
the Galactic foreground column towards NGC 55, 
$N_{H} = 1.72\times10^{20}~\rm cm^{-2}$ \citep{Dic90}. 
We tested to fix the column density at the Galactic value in fitting, but
the results were significantly worse for both models. We have also 
fitted both EPIC-PN and MOS spectra simultaneously using both models 
and the best-fit values are consistent with above quoted values.

The transient source was not detected in the 2001 {\it XMM-Newton} observation 
and we estimated the upper limit on the count rate 
using {\tt EREGIONANALYSE} task in SAS. After accounting for the background, 
we calculated a 90\% confidence upper limit of 
$< 1.5\times10^{-3}~\rm ct~s^{-1}$ on the count rate in 0.3--8 keV. 
We used the absorbed disk blackbody model to calculate the flux upper limit 
by assuming $kT_{in} = 0.7~\rm keV$ and $N_{H} = 0.3\times10^{22}~\rm cm^{-2}$.
The upper limit on the 0.3--8 keV flux is $< 1.4\times 10^{-14}~\rm erg~cm^{-2}~s^{-1}$ for 
this observation, which indicates that the flux from this source changed 
by a factor of $\sim$11. For the {\it Chandra} observations,
we analysed the ACIS data, and no sources were found
at the position of XMMU J001446.81-391123.48 in the 0.3--8\,keV or 3--8\,keV 
images.  Thus, we computed the 90\% confidence upper limits on the count rates, 
using {\tt APRATES} task in CIAO. The upper limits in the two epochs are 
$< 3.2\times10^{-4}~\rm$ and $< 1.1\times10^{-3}~\rm ct~s^{-1}$ respectively 
and the corresponding flux upper limits are $< 2.7\times10^{-15}$ and 
$< 8.2\times10^{-15}~\rm erg~cm^{-2}~s^{-1}$. The former value constrains
the source luminosity to be $< 10^{36}$\,erg\,s$^{-1}$.

\subsection{\swift{} XRT Data Analysis}

In the {\it Swift} XRT observations, using the {\tt XRTCENTROID} tool,
we derived the position of XMMU J001446.81-391123.48 
with a positional uncertainty of $\sim$ 5 arcsec, at 90\% confidence level. 
For each observation, we extracted source and background spectra from 
a circular region of radius 20 arcsec centred on the source position 
determined with {\tt XRTCENTROID}. The standard grade filtering of 0--12 was
used. The ancillary response files were created using 
the tool {\tt XRTMKARF} and appropriate response matrix files obtained 
from the HEASARC CALDB.

\begin{figure}
\centering
\includegraphics[width=8.5cm,height=8.5cm,angle=0]{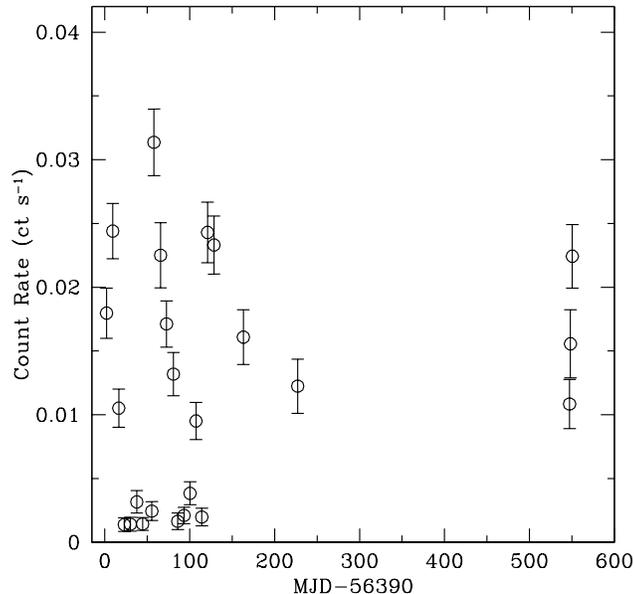}
\caption{The 0.3--8 keV \swift{} XRT count rate for the transient source as function of observation time.}
\label{lc}
\end{figure}

The Fig. \ref{lc} shows the {\it Swift} XRT light curve 
in the 0.3--8 keV energy range. From the light curve, it is clear that 
the source exhibited substantial variability and showed distinct intensity 
states. During 2013 April observations, the source is in 
the ``flaring state'' and then becomes fainter on $\sim \rm~day~20 $ in 
the light curve. The source remains in the ``faint state'' for about a month 
and then reaches the peak intensity ($\sim 0.03~\rm~ct~s^{-1}$) on 2013 
June 5. This intensity then decreases gradually and becomes fainter again 
after a month. The source reaches the peak intensity in $< 2.5$\,d 
and then declines by a factor of $\sim 19$ in $\sim 28$\,d, to reach 
the fainter state. Thus, the source appears to have a Fast Rise 
Exponential Decay \citep[FRED;][]{Che97} kind of phenomenon in the light curve. 
During 2013 July -- November observations, the source shows the same trend 
in the variations, but we could not follow up these variations because of 
the lack of monitoring. In addition, the source is in the rising stage 
during the latest \swift{} observations (2014 October). 

We investigated whether the observed intensity variation could be due 
to the change of the source position in the \swift{} CCD. We measured 
the flux of the source using the best-fit power law and disk blackbody 
models, which considered the effective area changes due to off-axis angle 
as well as the instrument response variations in the CCD. The estimated 
flux of the source also exhibits the variation pattern similar to that 
of the count rate. Thus we conclude that the intensity variation
came from the source, not due to instrumental artefacts.

\begin{figure}
\centering
\includegraphics[width=8.5cm,height=8.5cm,angle=0]{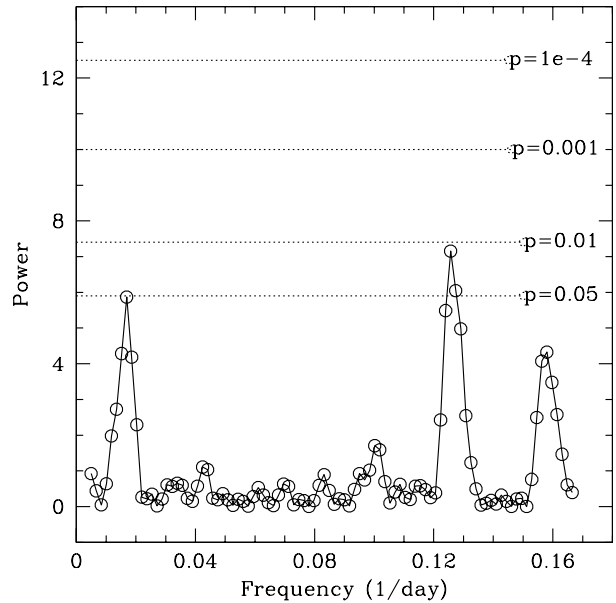}
\caption{The Lomb-Scargle periodogram for the \swift{} observations. The dotted lines represent different significance levels of detection.}
\label{lsp}
\end{figure}

In order to search for any modulation in the light curve, we used the Lomb-Scargle 
periodogram \citep[LSP;][]{Lom76,Sca82} and found a peak at $\sim 7.8$\,d 
in the LSP. However, the peak is not statistically significant 
($< 99\%$ level) with a maximum power of 7.1 (See Fig. \ref{lsp}).

Because of the low count rate, we co-added the \swift{} data to study 
the spectral behaviour of the source. We defined five count rate ranges 
as given in Table \ref{fit} by having roughly equal numbers of total counts 
(within a factor of 3). The spectra of each count rate group were added 
by using the {\tt FTOOL ADDSPEC} and the response files were weighted 
according to their counts. Since the statistical quality of the spectra
are low, they can be well fit by an absorbed power law or absorbed disk 
blackbody model. The absorption column density was fixed during 
the fit to the best-fit values of \xmm{} data. The best-fit spectral 
parameters are given in Table \ref{fit}.

While there is no substantial difference between the power law 
and disk blackbody fits for 4 co-added \swift{} spectra, for the {\it SWIFT4} 
spectrum, $\chi^2$/d.o.f are 49.9/43 and 65.2/43 from the power law 
and disk blackbody respectively, indicating that the former provides a better fit. 
We thus further studied the five individual spectra in the count rate 
range 0.02 -- 0.025. Fitting them with a power law and a disk blackbody, 4 out of 5 spectra in 
this count rate group were found to have similar spectral parameters (or 
within the uncertainty; $\Gamma\sim$ 2.5--3.0 or $kT_{in}\sim$ 0.7--0.9 keV), 
but one spectrum (ObsID : 00032619020) had $\Gamma\sim 2.21^{+0.41}_{-0.38}$ 
for the power law or $kT_{in}=1.23^{+0.50}_{-0.30}$\,keV for the disk 
blackbody. The $\chi^2$ values indicate that the power law is only slightly
more favoured, very similar to that for the 4 co-added spectra. We re-binned 
the spectra by removing the particular observation and fit 
the spectrum ({\it SWIFT4a} in Table~\ref{fit}) with both models. This fit results
also show a significant difference in $\chi^2$ between the two 
models ($\chi^2$/d.o.f are 31.1/33 and 45.6/33 for power law and disk 
blackbody respectively). Thus we conclude that the co-added spectra 
can not be used to determine the exact spectral parameters of the source, 
because they can appear to favour the power law when different individual 
spectra are added.

We also fitted the spectra by minimizing the Cash (C) statistic \citep{Cas79} 
for {\it SWIFT1}, {\it SWIFT2}, and {\it SWIFT5}. The C-stat values do 
not differ from the $\chi^2$ values, and the spectral parameters have 
negligible differences ($< 7\%$). It is also noted that the X-ray luminosity 
of the source is above $\sim 10^{38}~\rm erg~s^{-1}$ irrespective of 
the spectral models. 


\begin{table*}
\tabletypesize{\large}
\tablecolumns{10}
\setlength{\tabcolsep}{6.0pt}
\tablewidth{42pc}
	\caption{Spectral Fitting Parameters of XMMU J001446.81-391123.48.}
 	\begin{tabular}{@{}lccccccccr@{}}
	\hline
 & \colhead{Count} & & \colhead{$\rm PL^a$} & & & & \colhead{$\rm DISKBB^a$} & & \\
\colhead{Data}&\colhead{Rate$^b$}&\colhead{$n_{H}$$^c$}&\colhead{$\Gamma^d$}&\colhead{$L_{X}$$^e$}&\colhead{$\chi^2/d.o.f^f$}&\colhead{$n_{H}$$^c$}&\colhead{$kT_{in}$$^g$}&\colhead{$L_{X}$$^e$}&\colhead{$\chi^2/d.o.f^f$} 
\\    
\hline
{\it XMM1} & $-$ & $0.67^{+0.08}_{-0.07}$ & $3.17^{+0.22}_{-0.20}$ & $38.38^{+0.13}_{-0.11}$ & $74.2/71$ & $0.32^{+0.05}_{-0.04}$ & $0.70^{+0.06}_{-0.06}$ & $37.76^{+0.03}_{-0.03}$ & $70.45/71$ \\

{\it SWIFT1} & $0.00-0.01 (10)$ & $0.67(f)$ & $2.55^{+0.43}_{-0.39}$ & $38.19^{+0.14}_{-0.12}$ & $9.4/11$ & $0.32(f)$ & $0.98^{+0.35}_{-0.23}$ & $37.79^{+0.07}_{-0.08}$ & $11.7/11$ \\
{\it SWIFT2} & $0.01-0.015 (4)$ & $0.67(f)$ & $2.75^{+0.37}_{-0.34}$ & $38.85^{+0.13}_{-0.12}$ & $9.9/14$ & $0.32(f)$ & $0.88^{+0.21}_{-0.17}$ & $38.37^{+0.06}_{-0.06}$ & $12.5/14$ \\
{\it SWIFT3} & $0.015-0.02 (4)$ & $0.67(f)$ & $2.76^{+0.33}_{-0.28}$ & $39.04^{+0.12}_{-0.10}$ & $29.3/23$ & $0.32(f)$ & $0.94^{+0.19}_{-0.16}$ & $38.57^{+0.05}_{-0.05}$ & $34.3/23$ \\
{\it SWIFT4} & $0.02-0.025 (5)$ & $0.67(f)$ & $2.81^{+0.21}_{-0.20}$ & $39.16^{+0.08}_{-0.07}$ & $49.9/43$ & $0.32(f)$ & $0.84^{+0.11}_{-0.10}$ & $38.66^{+0.03}_{-0.04}$ & $65.2/43$ \\
{\it SWIFT4a} & $0.02-0.025 (4)$ & $0.67(f)$ & $2.89^{+0.25}_{-0.23}$ & $39.19^{+0.10}_{-0.09}$ & $31.1/33$ & $0.32(f)$ & $0.80^{+0.13}_{-0.11}$ & $38.67^{+0.04}_{-0.04}$ & $45.6/33$ \\
{\it SWIFT5} & $0.03-0.035 (1)$ & $0.67(f)$ & $2.65^{+0.51}_{-0.41}$ & $39.22^{+0.18}_{-0.14}$ & $15.9/11$ & $0.32(f)$ & $1.00^{+0.29}_{-0.23}$ & $38.80^{+0.07}_{-0.08}$ & $16.9/11$ \\

\hline
\end{tabular} 
\tablecomments {$^{a}$Spectral models used for fitting: PL - power law continuum; DISKBB - multi-colour disk blackbody. $^{b}$Count rate range used for the co-adding the \swift{} data and the no. of observations co-added are given in bracket. $^{c}$Absorption column density, including Galactic absorption, in $10^{22}~\rm cm^{-2}$. $^d$Power law index. $^e$Unabsorbed 0.3--8 keV X-ray luminosity in $\rm erg~s^{-1}$, calculated by assuming the distance of 1.78 Mpc \citep{Kar03}. $^f$The $\chi^2/d.o.f$ values for the model. $^g$Inner disk temperature in keV.}
\label{fit}
\end{table*}

\section{Discussion and Conclusion}

We serendipitously discovered a new transient source in NGC 55 using 
the archival {\it XMM-Newton} data obtained in 2010. 
The source was undetected in the {\it XMM-Newton} 2001 observation, 
and 2001 and 2004 \textit{Chandra} observations, whose deep sensitivities
indicate a flux change of at least two orders of magnitude from $<10^{36}$
to $\sim 10^{38}$\,erg\,s$^{-1}$. Thus, these observations establish 
that XMMU J001446.81-391123.48 is a new X-ray transient source in NGC 55. 

In considering the nature of XMMU J001446.81-391123.48, the X-ray colour 
classification scheme of \cite{Jen05} identifies this source as an X-ray
binary system. The high luminosities ($\sim \rm 10^{38}\rm erg~s^{-1}$)
during the outbursts are consistent with the luminosity range
of bright X-ray binaries, and also help to rule out the low-luminosity source 
classes, such as magnetic and non-magnetic Cataclysmic Variables \citep{Kuu06} 
and Magnetars \citep{Mer13}, as a possible candidate for this transient.    
Apart from the high luminosity, the source displayed a strong short-term 
X-ray variability in the 2010 observation, further supporting the X-ray binary nature. 

There is no statistically significant difference between the 
power law and disk blackbody fits for the \xmm{} spectrum. 
The best fit photon index ($\sim$ 3.2) is too high for the hard 
state \citep{Mcc06}; furthermore, the photon index is likely too high for 
the steep power law state also since the disk blackbody contribution makes 
the SPL appear hard in the observed energy range \citep{Bar14}. However, 
the best fit disk blackbody model is consistent with the thermally 
dominated spectrum \citep{Mcc06}, and we find this to be the most 
likely spectral state for the transient. 
Moreover, such a thermally dominated spectrum for an X-ray transient 
has never been seen in NS X-ray binaries, suggesting that the accretor 
in the transient is a black hole \citep{Don03}.

The follow-up studies with \swift{} XRT revealed the source's 
outburst activity and it is possibly a FRED phenomenon. During the 
2013 May \swift{} observations, the source had an outburst and reached 
the peak intensity, by having a factor $\sim 13$ flux increase, 
within $< 2.5$\,d. After the outburst, the flux decayed exponentially 
with an e-folding time of $> 22.8$\,d. This time scale is consistent with 
the e-folding time of black hole candidates in 
outburst \citep{Tan96, Che97, Yan14}. There are secondary peaks occurred 
prior (2013 April observations) and after the outburst 
(2013 August observations), but the peaks only reached 
a level of $\sim 78\%$ of the main outburst peak. In addition, the peak prior 
to the outburst decayed very rapidly compared to the main outburst decay. 
Thus, the source possibly showed a repeated outburst. The outburst recurrence 
period would be about a month, which is much smaller than the reported 
recurrence periods for the NS and BH systems \citep{Kuu97, Mas02}.

Our co-added \swift{} spectra only provide approximate spectral properties
for the source.  In the fainter 
state ({\it SWIFT1}), the spectrum is described by power law with index 
$\sim 2.5$ or disk blackbody with $kT_{in}\sim 1.0$\,keV. 
The source changed its luminosity by an order of magnitude 
and reached $\sim 2 \times 10^{39}~\rm erg~s^{-1}$ in the {\it SWIFT5} compared 
to {\it SWIFT1}, but not much change in the power law index or disk blackbody 
temperature. Although the large uncertainties on the parameters do not allow 
us to draw a clear conclusion, the results suggest that the source changed its 
luminosity by an order of magnitude without change in its state. 
In the observations, the X-ray luminosity of XMMU J001446.81-391123.48 
is $\sim 30-400\%$ of the Eddington limit for a canonical 
$\rm 1.4\ \Msun$ NS, again suggesting that the primary is more likely a BH. 

Finally, the transient source is located in the bar region which is displaced 
$\sim$ 3 arcmin from the geometrical centre of NGC 55 \citep{Rob66}. The Very 
Large Array observations at 6 cm and 21 cm revealed that the radio continuum 
emission is concentrated on the bar region \citep{Hum86}. Moreover, the 6 cm 
radio emission is dominated by a triple source and this triple coincides with 
discrete $\rm H_ {I}$ and $\rm H_{\alpha}$ emission. Also, the on-going star 
formation rate (SFR) in the bar region is consistent with the global SFR of 
NGC 55 \citep[$\sim \rm 0.22~\Msun yr^{-1}$;][]{Eng04} and suggests that the 
bar region is a young star formation complex with an age of $< \rm2~Myr$. 
Thus, the transient source could be a young stellar system in the bar region 
of NGC 55. 
 
In summary, XMMU J001446.81-391123.48 is a new X-ray transient in the young 
stellar region of NGC 55, possibly being a black hole X-ray binary. 
The repeated outbursts and prominent radio emission from
the young stellar region made this source as a good candidate for 
the further follow-up 
studies at X-ray and radio wavelengths. Such multi-wavelength studies can 
shed further light on the nature of XMMU J001446.81-391123.48.

\section*{Acknowledgements}

We thank referee for constructive suggestions, and
thank Ranjeev Misra and FuGuo Xie for useful comments 
on the manuscript. This work has made use of data obtained from the 
High Energy Astrophysics Science Archive Research Center (HEASARC), 
provided by NASA's Goddard Space Flight Center. This research was funded 
by Chinese Academy of Sciences President's International Fellowship 
Initiative (Grant No. 2015PM059), and was supported in part by the 
Strategic Priority Research Program ``The Emergence of Cosmological 
Structures" of the Chinese Academy of Sciences (Grant No. XDB09000000). 
ZW is a Research Fellow of the One-Hundred-Talents project of Chinese 
Academy of Sciences.


\end{document}